\documentclass{PoS}

\usepackage{aas_macros}

\usepackage[T1]{fontenc} 

\usepackage{bbm}
\usepackage{slashed}
\usepackage{color}

\usepackage{epsfig,graphicx,bm,color} 
\usepackage{dcolumn}
\usepackage{bm}
\usepackage{color}
\usepackage{soul} 

\usepackage{aas_macros}
\usepackage{tablefootnote}
\usepackage[displaymath]{lineno}
\usepackage{subfigure}
\usepackage{multirow}
\usepackage{amsmath}

\def\rpv{{R}_{p} \hspace{-0.4cm}\slash\hspace{0.2cm}}
\def\rp{{R}_{p}}

\def\lsim{\raise0.3ex\hbox{$\;<$\kern-0.75em\raise-1.1ex\hbox{$\sim\;$}}}
\def\gsim{\raise0.3ex\hbox{$\;>$\kern-0.75em\raise-1.1ex\hbox{$\sim\;$}}}

\def    \beq            {\begin{equation}}
\def    \eeq            {\end{equation}}
\def    \bea           {\begin{eqnarray}}
\def    \eea           {\end{eqnarray}}

\def \mn{\mu\nu{\rm SSM}}
\def\n{\widetilde\chi^0}

\def\g2{{\rm GeV}^2}

\def\n{\widetilde\chi^0}

\def\sw2{sin^2 \theta_w}

\def\a^tau{\alpha_{\tau}}

\def\beq{\begin{equation}}
\def\eeq{\end{equation}}
\def\beqa{\begin{eqnarray}}
\def\eeqa{\end{eqnarray}}

\newcommand{\newc}{\newcommand}
\newc\BR{BR}
\newc{\akappa}{A_{\kappa} }
\newc\deltagmtwo{\delta (g-2)_{\mu}} 
\newc\deltaamu{\Delta a_{\mu}}

\def\anti{\overline}

\def\rpv{{R}_{p} \hspace{-0.4cm}\slash\hspace{0.2cm}}

\newc{\haa}{BR\(h_1\to a_1 a_1\)}
\newc{\abb}{BR\(a_1\to b\anti{b}\)}
\newc{\hbb}{BR\(h_1\to b\anti{b}\)}
\newc{\Fermi}{\textit{Fermi}-}
\newc{\abund}{\Omega h^2}
\newc\bsgamma{b\rightarrow s \gamma }
\newc\bxsgamma{\overline{B}\rightarrow X_{s}\gamma}
\newc\brbsgamma{\BR(\overline{B}\rightarrow X_s\gamma)}

\title{Searching for SUSY and decaying gravitino DM
\\
at the LHC and Fermi-LAT with the $\mu\nu$SSM}

\ShortTitle{
SUSY and decaying gravitino DM at the LHC and Fermi-LAT with the $\mu\nu$SSM}

\author{\speaker{CARLOS MU\~NOZ}%
\\
        Departamento de F\'isica Te\'orica,
Universidad Aut\'onoma de Madrid (UAM), 28049 Spain\\
Instituto de F\'isica Te\'orica UAM--CSIC, Campus de Cantoblanco UAM, 28049 Spain
\\
        E-mail: \email{c.munoz@uam.es}}

\abstract{
The `$\mu$ from $\nu$' supersymmetric standard model ($\mu\nu$SSM) solves the $\mu$ problem of supersymmetric models and reproduces neutrino data, simply using couplings with the three families of right-handed neutrinos $\nu$'s. Novel signatures of supersymmetry at the LHC are expected through these new states, and couplings breaking $R$ parity.
All supersymmetric particles are potential candidates for the lightest one, which is not stable 
leading to prompt or displaced vertices and producing final states with multi-leptons/taus/jets/photons and missing energy. 
Besides, a decaying gravitino turns out to be an interesting candidate for dark matter. It can be searched through gamma-ray observations, such as those of the Fermi Large Area Telescope.
The latter, depending on the region of the parameter space of the model, already imposes an upper bound on the gravitino mass of the order of $5-20$ GeV and a lower bound on the lifetime of about $10^{25-28}$ s. 
}

\FullConference{11th International Workshop Dark Side of the Universe 2015\\
		 14-18 December 2015\\
		 Yukawa Institute for Theoretical Physics, Kyoto University Japan}

\begin{document}

\section{Introduction}

Supersymmetry (SUSY) is still the most compelling theory for physics beyond the standard model. 
SUSY not only solves several important theoretical problems of the standard model, such as the gauge hierarchy problem and others, but also has spectacular experimental implications. As is well known, the spectrum of elementary particles is doubled with masses of about 1 TeV, thus even
the simplest SUSY model, the minimal supersymmetric standard model (MSSM, see Ref.~\cite{Martin:1997ns} for a review), predicts a rich phenomenology.
However, the LHC started operations several years ago and, with Run 1 already finished, SUSY has not been discovered yet. Because of this, it has been raised the question of whether SUSY is still alive. The question is fair of course, but in our opinion the answer is yes, and we think that there are several arguments in favor of this answer. Here there are some of them:

$\bullet$ The lower bounds on SUSY particle (sparticle) masses are smaller than 1 TeV or about that number, depending on the sparticle analyzed. Thus the SUSY masses are still reasonable, and in that sense we can keep in mind what happened before the discovery of the Higgs boson.

$\bullet$ Because of the complicated parameter space of SUSY, experimentalists use in their analyses simplified models that do not cover the full MSSM. For example, branching-ratio variations are not considered in much detail, and other assumptions are also made.

$\bullet$ Run 2 is going on, and for the moment with a low luminosity of about 20 fb$^{-1}$.
Therefore, to (be prepared) wait for the results with higher luminosity seems to be a sensible strategy, since 100 fb$^{-1}$ are expected for the end of the Run 2.
 
$\bullet$ Most searches at the LHC assume $R$-parity conservation ($\rp$),
with the lightest supersymmetric particle (LSP) stable, requiring therefore missing energy in the final state to claim for SUSY detection. But, if $R$ parity is violated 
($\rpv$), 
sparticles can decay to standard model particles, and the bounds on their masses can become significantly weaker.

Nevertheless, despite all these arguments, it is also honest to recognize that SUSY has its own theoretical problems in its formulation at low energy. and, in particular, a crucial one is the so-called
$\mu$ problem \cite{Kim:1983dt}. In the superpotential of the MSSM
\begin{equation}
W = 
\epsilon_{ab} \left(
Y_{u_{ij}} \, \hat H_u^b\, \hat Q^a_i \, \hat u_j^c +
Y_{d_{ij}} \, \hat H_d^a\, \hat Q^b_i \, \hat d_j^c +
Y_{e_{ij}} \, \hat H_d^a\, \hat L^b_i \, \hat e_j^c \right)
-
\epsilon{_{ab}} \mu \,\hat H_d^a \hat H_u^b
\ ,
\label{superpotential1}
\end{equation}
the presence of the mass parameter $\mu$ is necessary, for example to generate Higgsino masses given the current experimental lower bound of about 100 GeV on chargino masses.
In the presence of a high-energy theory like a grand unified theory (GUT) or a string theory, with a typical scale of the order of $10^{16}$ GeV or larger, and/or a gravitational theory at the Planck scale, one should be able to explain how to obtain a mass parameter in the superpotential of the order of the electroweak (EW) scale. 
The MSSM does not solve the $\mu$ problem. One takes for granted that the 
$\mu$ term is there and that is of the order of the EW scale, and that's it.
In this sense, the MSSM is a kind of effective theory.

Another theoretical problem of SUSY is to be able to build a model solving the $\nu$ problem: 
how to reproduce neutrino data~\cite{Tortola:2012te}, i.e. masses and mixing angles.
Let us emphasize in this sense that in the MSSM, by construction, neutrinos are massless.

The 
`$\mu$ from $\nu$' supersymmetric standard model
($\mu\nu$SSM~\cite{LopezFogliani:2005yw,Escudero:2008jg}, see Refs.~\cite{Munoz:2009an,LopezFogliani:2010bf} for reviews),
includes new couplings with the three families of right-handed (RH) neutrino superfields 
($\hat \nu^c_i$ with $i=1,2,3$) in the superpontential
in order to solve the 
$\mu$-problem, while simultaneously explains the origin of neutrino masses.
The $SU(3)_c\times SU(2)_L\times U(1)_Y$ invariant couplings
$\lambda_i \hat \nu^c_i \hat H_d\hat H_u$ generate an
effective  $\mu$ term through RH sneutrino vacuum expectation values (VEVs),
$\langle \tilde \nu^c_i \rangle \equiv v_{\nu^c_i}$, after the successful 
electroweak symmetry breaking (EWSB): $\mu^{eff}=\lambda_i v_{\nu^c_i}$.
In addition, other gauge invariant couplings $\frac{1}{3}\kappa{_{ijk}} \hat \nu^c_i\hat \nu^c_j\hat \nu^c_k$
generate effective Majorana masses for the RH neutrinos,
$M_{ij}^{eff}=2\kappa_{ijk}v_{\nu^c_k}$, contributing to a generalized EW-scale seesaw mechanism
which can reproduce the observed neutrino masses and mixing angles.
We will review this solution to the $\mu$ problem and neutrino physics in 
Section~\ref{model}.

Since sparticles do not appear in pairs in these couplings that solve the $\mu$ and
$\nu$ problems, we say in the usual language that they produce explicit $\rpv$.
This implies that the phenomenology of the 
$\mu\nu$SSM is very different from the one of the MSSM or the next-to-MSSM (NMSSM, see Ref.~\cite{Ellwanger:2009dp} 
for a review).  
We will briefly review the phenomenology of the $\mn$ at the LHC in 
Section~\ref{lhc}. There we will see that since 
the LSP is not stable because of $\rpv$,
it decays leading to
prompt or displaced vertices, and producing final states with multi-leptons/taus/jets/photons and missing energy.

On the other hand, the usual sparticle candidates for the dark matter (DM) of the Universe in the case of $\rp$, the neutralino or the RH sneutrino, 
have very short lifetimes in $\rpv$
models, and therefore can no longer be used.
Nevertheless, the
gravitino 
can still be a candidate for DM
since its lifetime is typically very long, being
suppressed both by
the gravitational interaction and by the small $\rpv$
couplings.
In Section~\ref{darkmatter}, we will discuss the feasibility of gravitino DM in the $\mu\nu$SSM, as well as 
its possible detection in gamma-ray satellite experiments such as the
Fermi Large Area Telescope (LAT).
Our conclusions are left for Section~\ref{conclusions}.

\section{The $\mu\nu$SSM}
\label{model}

The superpotential of the $\mu\nu$SSM 
contains in addition to the MSSM
Yukawas for quarks and charged leptons,
Yukawas for neutrinos and the two couplings discussed in the introduction that
generate the effective $\mu$ term and Majorana masses \cite{LopezFogliani:2005yw,Escudero:2008jg}:
\begin{eqnarray}
W & = &
\ \epsilon_{ab} \left(
Y_{u_{ij}} \, \hat H_u^b\, \hat Q^a_i \, \hat u_j^c +
Y_{d_{ij}} \, \hat H_d^a\, \hat Q^b_i \, \hat d_j^c +
Y_{e_{ij}} \, \hat H_d^a\, \hat L^b_i \, \hat e_j^c +
Y_{\nu_{ij}} \, \hat H_u^b\, \hat L^a_i \, \hat \nu^c_j 
\right)
\nonumber\\
& - &
\epsilon{_{ab}} \lambda_{i} \, \hat \nu^c_i\,\hat H_d^a \hat H_u^b
+
\frac{1}{3}
\kappa{_{ijk}} 
\hat \nu^c_i\hat \nu^c_j\hat \nu^c_k \ .
\label{superpotential}
\end{eqnarray}
Notice that 
in the limit $Y_{\nu_{ij}} \to 0$, $\hat \nu^c_i$ can be identified as
pure singlet superfields without lepton number, similar to the case of the NMSSM,
where one singlet is added to the spectrum and there is $\rp$. Thus $\rpv$ in the $\mu\nu$SSM is determined by the values of the neutrino Yukawa couplings, and as a consequence is going to be small.

\subsection{The Solution to the $\mu$ Problem}

To confirm that the $5^{\rm th}$ term in the superpotential generates dynamically the $\mu$ term
(and the $6^{\rm th}$ term the Majorana masses for neutrinos),
as discussed in the Introduction, we have to probe that 
the VEVs of the RH sneutrinos
are naturally of the order of the EWSB scale.
Since only dimensionless trilinear couplings are present in 
(\ref{superpotential}), the EWSB
is determined by the usual soft SUSY-breaking terms of the scalar potential. 
Thus it is remarkable that all known particle physics phenomenology can be reproduced in the $\mu\nu$SSM with one scale, avoiding the introduction of {\it ad-hoc} high-energy scales like e.g. in the GUT-scale seesaw.

To carry out the minimization, let us remember that 
in addition to the soft terms the tree-level neutral scalar potential
receives the $D$ and $F$ term contributions 
that can be found in Refs.~\cite{LopezFogliani:2005yw, Escudero:2008jg}.
With the choice of CP conservation,\footnote{$\mu\nu$SSM with spontaneous 
CP violation was studied in Ref.~\cite{Fidalgo:2009dm}.} after the EWSB the neutral scalars
develop in general the following real VEVs:
\begin{equation}
\langle H_d^0 \rangle = v_d\ , \, \quad \langle H_u^0 \rangle = v_u\ , \,
\quad \langle \widetilde \nu_i \rangle = v_{\nu_i}\ , \,  \quad
\langle \widetilde \nu_i^c \rangle = v_{\nu^c_i}\ ,
\label{vevs}
\end{equation}
%
where in addition to the usual VEVs of the MSSM Higgses, $H_u^0$ and $H_d^0$, the new couplings generate VEVs for left-handed (LH) sneutrinos, $\widetilde \nu_i$, as well as for the RH sneutrinos, $\widetilde \nu_i^c$.
The eight minimization conditions 
can be written as
{\small
\begin{eqnarray}
 m_{H_{d}}^{2}
& =& -\frac{1}{4}G^2\left(v_{\nu_{i}}v_{\nu_{i}}+v_{d}^{2}-v_{u}^{2}\right)
 - \lambda_i\lambda_{j}v_{\nu_i^c} v_{\nu_j^c} 
 - \lambda_{i}\lambda_{i}
v_{u}^{2}
\nonumber\\ 
  &&
   +v_{\nu_i^c}\tan\beta\left(a_{\lambda_i} 
   + \lambda_{j}\kappa_{ijk}
v_{\nu_k^c} 
\right)
   + Y_{\nu_{ij}} \frac{v_{\nu_{i}}}{v_d}
\left(\lambda_k v_{\nu_k^c} v_{\nu_j^c} +
\lambda_{j} 
v_{u}^2\right)
\ , 
   \label{tadpoles1}
\\ 
\nonumber
\\
 m_{H_{u}}^{2}
& =& \frac{1}{4}G^2\left(v_{\nu_{i}}v_{\nu_{i}}+v_{d}^{2}-v_{u}^{2}\right)
 - \lambda_{i}\lambda_j
v_{\nu_{i}^c}v_{\nu_{j}^c}
-\lambda_{j}\lambda_{j}v_{d}^2
\nonumber\\
&&
+ 2\lambda_j Y_{\nu_{ij}}v_{\nu_{i}}v_{d} 
   - Y_{\nu_{ij}}Y_{\nu_{ik}}
v_{\nu_{k}^c}v_{\nu_{j}^c}
- Y_{\nu_{ij}}Y_{\nu_{kj}}v_{\nu_{i}}v_{\nu_{k}} 
   \nonumber \\ 
&&  
+v_{\nu^c_i} \frac{1}{\mathrm{tan} \beta}\left(a_{\lambda_i}
+\lambda_{j}\kappa_{ijk}v_{\nu^c_k}\right)
-\frac{v_{\nu_i}}{v_u}\left(a_{\nu_{ij}}v_{\nu^c_j}+
Y_{\nu_{ij}}\kappa_{ljk}v_{\nu^c_l}v_{\nu^c_k}\right) 
\ , 
   \label{tadpoles2}
\\ 
\nonumber\\
  m^2_{\widetilde{\nu}_{ij}^{c}}v_{\nu_{j}^{c}} &=&    
   -a_{\nu_{ji}}v_{\nu_{j}}v_{u}
   + a_{\lambda_i}v_u v_d - a_{\kappa_{ijk}}v_{\nu_{j}^c}v_{\nu_{k}^c} 
   - \lambda_i\lambda_{j}\left(v_{u}^{2}+v_{d}^{2}\right)v_{\nu_{j}^c} 
   + 2\lambda_{j}\kappa_{ijk}v_{d}v_{u}v_{\nu_{k}^c}
   \nonumber \\ 
  &&
   - 2\kappa_{lim}\kappa_{ljk}v_{\nu_{m}^c}v_{\nu_{j}^c}v_{\nu_{k}^c}
   + Y_{\nu_{ji}}\lambda_{k}v_{\nu_{j}}v_{\nu_{k}^c}v_{d}
   + Y_{\nu_{kj}}\lambda_{i}v_{d}v_{\nu_{k}}v_{\nu_{j}^c}
- 2 Y_{\nu_{jk}}\kappa_{ikl}v_{u}v_{\nu_{j}}v_{\nu_{l}^c}
   \nonumber \\ 
  &&
    - Y_{\nu_{ji}}Y_{\nu_{lk}}v_{\nu_{j}}v_{\nu_{l}}v_{\nu_{k}^c}
    - Y_{\nu_{ki}}Y_{\nu_{kj}}v_{u}^{2}v_{\nu_{j}^c}
\ ,  
     \label{tadpoles3}
     \\
    \nonumber\\
  m^2_{\widetilde{L}_{ij}}v_{\nu_{j}}& =&  -\frac{1}{4}G^2\left(v_{\nu_{j}}v_{\nu_{j}}+v_{d}^{2}-v_{u}^{2}\right)v_{\nu_{i}}
   -a_{\nu_{ij}}v_{u}v_{\nu_{j}^c} 
   + Y_{\nu_{ij}}\lambda_{k}v_{d}v_{\nu_{j}^c}v_{\nu_{k}^c}
     + Y_{\nu_{ij}}\lambda_{j}v_{u}^{2}v_{d}
     \nonumber \\
 &&
     - Y_{\nu_{il}}\kappa_{ljk}v_{u}v_{\nu_{j}^c}v_{\nu_{k}^c}
 - Y_{\nu_{ij}} Y_{\nu_{lk}}v_{\nu_{l}}v_{\nu_{j}^c}v_{\nu_{k}^c}
   - Y_{\nu_{ik}}Y_{\nu_{jk}}v_{u}^{2}v_{\nu_{j}^c}
\ ,
\label{tadpoles4}
\end{eqnarray}
}
where the low-energy soft masses 
$m_{H_{d}}^{2}$, $m_{H_{u}}^{2}$, $m^2_{\widetilde{\nu}_{ij}^{c}}$ and
$m^2_{\widetilde{L}_{ij}}$ are calculated as functions of the VEVs $v_d$, $v_u$, $v_{\nu^c_i}$,
$v_{\nu_i}$. Besides, inspired by the structure of supergravity, the soft trilinear parameters are taken directly proportional to the couplings,
$a_{\lambda_i}= A_{\lambda_i}\lambda_i$, 
$a_{\kappa_{ijk}}= A_{\kappa_{ijk}} \kappa_{ijk}$,
$a_{\nu_{ij}}= A_{\nu_{ij}} Y_{\nu_{ij}}$, etc., where
the summation convention on repeated indices does not apply for these particular formulas.

As can be easily seen from Eq.~(\ref{tadpoles3}),
the VEVs of the RH sneutrinos, $v_{\nu_{j}^{c}}$, 
are naturally of the order of the EWSB scale, confirming that the solution to the $\mu$ problem works.

\subsection{The solution to the $\nu$ problem}
\label{nuproblem}

Using the same argument as above, we trivially confirm that the $6^{\rm th}$ term in the superpotential (\ref{superpotential})
generates the effective Majorana masses for RH neutrinos, as discussed in the Introduction.
Thus we can implement naturally an EW-scale seesaw
in the $\mu\nu$SSM, asking for 
neutrino Yukawa 
couplings of the order of the electron Yukawa coupling or smaller,
$Y_{\nu_{ij}} \sim 10^{-6} - 10^{-7}$ \cite{LopezFogliani:2005yw,
Escudero:2008jg,Ghosh:2008yh,Bartl:2009an,Fidalgo:2009dm,Ghosh:2010zi,LopezFogliani:2010bf,Ghosh:2010ig},
i.e. we work with Dirac masses for neutrinos,
$m_D\sim Y_{\nu}v_u\lsim 10^{-4}$ GeV.
On the other hand, the VEVs 
of the LH sneutrinos,
$v_{\nu_i}$, are much smaller than the other VEVs (\ref{vevs}) in the $\mu\nu$SSM.
Notice in this respect that in Eq.~(\ref{tadpoles4}),
$v_{\nu}\to 0$ as $Y_{\nu}\to 0$.
It is then easy to estimate the values of these VEVs as $v_{\nu}\lsim m_D$~\cite{LopezFogliani:2005yw}.

As is well known, the couplings and Higgs VEVs present in the MSSM (determined by the superpotential (\ref{superpotential1})) generate the mixing of neutral gauginos and Higgsinos, where the eigenstates are the so-called neutralinos.
A similar situation occurs in the $\mu\nu$SSM, however in this model there are new couplings and VEVs (see Eqs.~(\ref{superpotential}) and (\ref{vevs})), implying larger mass matrices than those of the MSSM or NMSSM.
In particular, in the case of the neutralinos, they turn out to be also mixed 
with the LH and RH neutrinos. Besides, we have seen before that Majorana masses for RH neutrinos are generated dynamically, thus they will behave as the singlino components of the neutralinos.
Altogether, in a basis where
${\chi^{0}}^T=(\tilde{B^{0}},
\tilde{W^{0}},\tilde{H_{d}},\tilde{H_{u}},\nu_{R_i},\nu_{L_i})$,
one obtains the following $10\times 10$ neutral fermion
(neutralino-neutrino) mass
matrix \cite{LopezFogliani:2005yw,Escudero:2008jg}:
%
\begin{equation}
{\mathcal M}_n=
\left(
\begin{array}{cc}
M & m\\
m^{T} & 0_{3\times3}
\end{array}
\right)\ ,
\label{mixing}
\end{equation}
with

\begin{equation}
M=
\left(
\begin{array}{ccccccc}
M_{1} & 0 & -A v_{d} & A v_{u} & 0 & 0 & 0\\
0 & M_{2} &B v_{d} & -B v_{u} & 0 & 0 & 0\\
-A v_{d} & B v_{d} & 0 & -\lambda_{i}v_{\nu^c_{i}} & -\lambda_{1}v_{u} & -\lambda_{2}v_{u} & -\lambda_{3}v_{u}\\
A v_{u} & -B v_{u} & \: \: -\lambda_{i}v_{\nu^c_{i}} & 0 & -\lambda_{1}v_{d}+Y_{\nu_{i1}}v_{\nu_{i}} & -\lambda_{2}v_{d}+Y_{\nu_{i2}}v_{\nu_{i}} & -\lambda_{3}v_{d}+Y_{\nu_{i3}}v_{\nu_{i}}\\
0 & 0 &  -\lambda_{1}v_{u} & \: \:-\lambda_{1}v_{d}+Y_{\nu_{i1}}v_{\nu_{i}} & 2\kappa_{11j}v_{\nu^c_{j}} & 2\kappa_{12j}v_{\nu^c_{j}} & 2\kappa_{13j}v_{\nu^c_{j}}\\
0 & 0 & -\lambda_{2}v_{u} &  \: \: -\lambda_{2}v_{d}+Y_{\nu_{i2}}v_{\nu_{i}} & 2\kappa_{21j}v_{\nu^c_{j}} & 2\kappa_{22j}v_{\nu^c_{j}} & 2\kappa_{23j}v_{\nu^c_{j}}\\
0 & 0 & -\lambda_{3}v_{u} & \: \:-\lambda_{3}v_{d}+Y_{\nu_{i3}}v_{\nu_{i}} & 2\kappa_{31j}v_{\nu^c_{j}} & 2\kappa_{32j}v_{\nu^c_{j}} & 2\kappa_{33j}v_{\nu^c_{j}}
\end{array}
\right)\ ,
\label{neumatrix}
\end{equation}
where $A\equiv\frac{G}{\sqrt{2}} \sin\theta_W$, $B\equiv\frac{G}{\sqrt{2}} \cos\theta_W$,
with $G^2\equiv g_{1}^{2}+g_{2}^{2}$,
and
\begin{equation}
m^{T}=
\left(\begin{array}{ccccccc}
-\frac{g_{1}}{\sqrt{2}}v_{\nu_{1}} \: & \: \frac{g_{2}}{\sqrt{2}}v_{\nu_{1}} & \: 0 & \: Y_{\nu_{1i}}v_{\nu^c_{i}} & \: Y_{\nu_{11}}v_{u} & \: Y_{\nu_{12}}v_{u} & \: Y_{\nu_{13}}v_{u}\\
\: -\frac{g_{1}}{\sqrt{2}}v_{\nu_{2}} & \: \frac{g_{2}}{\sqrt{2}}v_{\nu_{2}} & \: 0 & \: Y_{\nu_{2i}}v_{\nu^c_{i}} & \: Y_{\nu_{21}}v_{u} & \: Y_{\nu_{22}}v_{u} & \: Y_{\nu_{23}}v_{u}\\
\: -\frac{g_{1}}{\sqrt{2}}v_{\nu_{3}}\: & \: \frac{g_{2}}{\sqrt{2}}v_{\nu_{3}} & \: 0 & \: Y_{\nu_{3i}}v_{\nu^c_{i}} & \: Y_{\nu_{31}}v_{u} & \: Y_{\nu_{32}}v_{u} & \: Y_{\nu_{33}}v_{u}\end{array}\right)\ .
\label{mixing3}
\end{equation}
The structure of this mass matrix is that of a 
generalized EW-scale seesaw, since it involves not only the RH neutrinos but also the neutralinos. 
Because of this structure, data on neutrino physics can easily be 
reproduced at tree level \cite{LopezFogliani:2005yw,
Escudero:2008jg,Ghosh:2008yh,Fidalgo:2009dm,Ghosh:2010zi}, even with diagonal 
Yukawa couplings $Y_{\nu_i}$ \cite{Ghosh:2008yh,Fidalgo:2009dm}.
Qualitatively, we can understand this in the following way. First of all, neutrino masses are going to be very small since
the entries of the matrix $M$ are much larger than the ones of the matrix $m$. Notice in this sense that the entries of $M$ are of the order of the EW scale, whereas the ones in $m$ are of the order 
of the Dirac masses for neutrinos \cite{LopezFogliani:2005yw,
Escudero:2008jg}. 
Second, from the above matrices, in the limit of large $\tan\beta$ 
(where $\tan\beta\equiv v_u/v_d$) one can obtain a simplified formula for the effective neutrino mixing mass matrix~\cite{Fidalgo:2009dm}:
\begin{eqnarray}
(m^{eff}_{\nu})_{ij} 
\simeq \frac{Y_{\nu_i}Y_{\nu_j}v_u^2}
{6 \kappa v_{\nu^c}}
                   (1-3 \delta_{ij})-\frac{v_{\nu_i} v_{\nu_j}}{2M}  \ ,
\label{Limit no mixing Higgsinos gauginos}
\end{eqnarray}
where $\kappa_{iii}\equiv \kappa_{i} \equiv \kappa$ and vanishing otherwise,
$v_{\nu^c_i}\equiv v_{\nu^c}$, and
${M} \equiv \frac{M_1 M_2}{g_1^2 M_2 + g_2^2 M_1}$. 
Using this approximate formula it is easy
to understand how diagonal Yukawas, $Y_{\nu_{ii}}=Y_{\nu_i}$ and vanishing otherwise, can give rise to off-diagonal entries in the mass matrix. One of the key points is the extra contribution given by the first term of 
Eq.~(\ref{Limit no mixing Higgsinos gauginos}) with respect to the ordinary seesaw where it is absent. Another extra contribution to the off-diagonal entries is the last term, which is generated through the mixing of LH neutrinos with gauginos.




In a sense, all these arguments give an answer to the question why the mixing angles are so different in the quark and lepton sectors: because no generalized seesaw exists for the quarks.

\section{LHC Phenomenology}
\label{lhc}

We have already mentioned in the previous section that mass matrices in the $\mn$ are larger than in the
MSSM or NMSSM, because of the presence of new couplings and VEVs. We also discussed the case of the 
$10\times 10$ neutral fermion (neutralino-neutrino) mass matrix.
For the rest of the mass matrices, a similar situation occurs and new mixing of states are induced~\cite{LopezFogliani:2005yw, Escudero:2008jg}.
Summarizing, there are 
five charged fermions (charginos-charged leptons),
seven CP-odd and eight CP-even neutral scalars (Higgses-sneutrinos),
and seven charged scalars (charged Higgses-sleptons).
As a consequence, the phenomenology of the $\mu\nu$SSM is very different from the one of the MSSM or NMSSM.

Needless to mention, as in $\rpv$ models 
the LSP 
is no longer stable, and
therefore not all SUSY chains must yield
missing energy events at colliders.
In particular, in the $\mu\nu$SSM
the LSP decays leading to
prompt or displaced vertices, depending on the value of the couplings, and producing final states with multi-leptons/taus/jets/photons and missing energy.
This unusual phenomenology was explored first in 
Refs.~\cite{Bartl:2009an,Bandyopadhyay:2010cu,Fidalgo:2011ky,Ghosh:2012pq}, discussing the decay properties of the LSP assumed to be the lightest neutralino, as well as novel Higgs decays. Further, detailed collider analyses
for a Higgs-like scalar decaying into a pair of neutralinos was also discussed
in Refs. \cite{Bandyopadhyay:2010cu,Ghosh:2012pq}, provided that these states lie below in the mass spectrum. 
More recently, this issue was revisited and, under the same assumption, a Higgs-like scalar decaying to a pair of scalars/pseudoscalars was also considered \cite{Ghosh:2014ida}.
The case of non-standard on-shell decays of $W^\pm$ and $Z$ bosons to light singlet-like scalar(s), pseudoscalar(s) and neutralinos(s) was studied in Ref.~\cite{Ghosh:2014rha}.

On the other hand, all sparticles are potential LSP's in $\rpv$ models, since the problem of stable charged particles as DM is not present. So to study the whole potential phenomenology of the $\mu\nu$SSM at the LHC, we should be prepared to analyze systematically not only the usual lightest neutralino as the LSP, but also 
the lightest stau, squark, chargino, and sneutrino as LSP's with a wide range of masses. In a first detailed analysis \cite{sneutrino} we have concentrated in the LH sneutrino as the LSP. We have shown that
for a sneutrino mass in the range about $95-145$ GeV, 
a diphoton 
signal plus leptons, or plus missing transverse energy (from neutrinos),
is observable at the LHC, even at the current Run 2 with 100 fb$^{-1}$  of luminosity.
The dominant sneutrino pair production channels are 
the direct production via a $Z$ boson, or through a $W^{\pm}$ decaying into a sneutrino and a LH charged slepton next-to-LSP, with the latter decaying into another sneutrino plus a very soft $W^{\pm}$. 
We think that these signals (where one of the sneutrinos decays in a way not very different from the Higgs) are worthy of attention by our experimental colleagues.

\section{Gravitino Dark Matter}
\label{darkmatter}

\begin{figure}[t]
 \begin{center}
 \includegraphics[
width=0.35
\textwidth]
{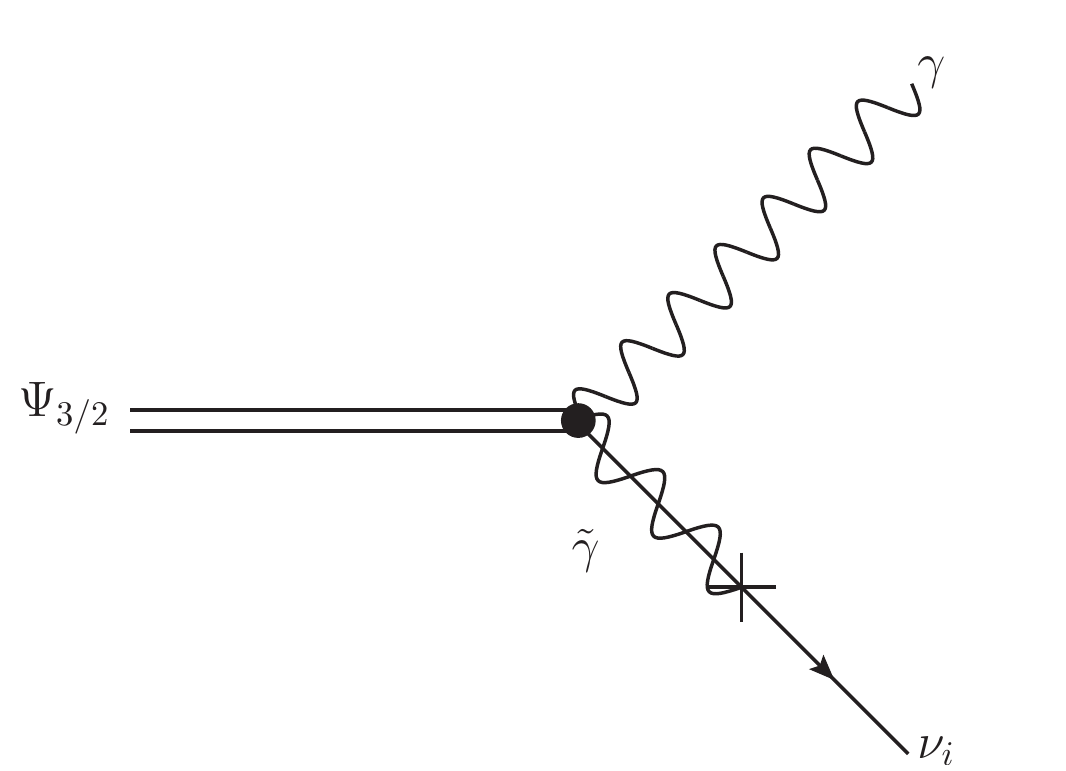}
\caption{Tree-level diagram for the two-body decay of a gravitino into a photon and a neutrino, via photino-neutrino mixing.}
    \label{decay}
 \end{center}
\end{figure}

As already mentioned in the Introduction, the gravitino is an interesting candidate for DM in $\rpv$ models. This occurs when it becomes the LSP.
The gravitino has an interaction term in the supergravity Lagrangian
with 
the photon and the photino. Since
the photino and the LH neutrinos are mixed in the neutral fermion mass matrix due to the $\rpv$, as discussed in Eq.~(\ref{mixing3}),
the gravitino will be able to decay 
into a photon and a neutrino, as shown in Fig.~\ref{decay}. Nevertheless, this decay is suppressed both by
the gravitational interaction and by the small $\rpv$ coupling, making
the gravitino lifetime much longer than the age of the Universe \cite{Takayama:2000uz}.
From the supergravity Lagrangian one obtains
\begin{equation}
\Gamma\left(\Psi_{3/2}
\to \sum_i \gamma \nu_i\right)\simeq  
\frac{1}{64 \pi} |U_{\widetilde{\gamma} \nu}|^2\,\frac{m^3_{3/2}}{M_P^2}\,, 
\label{c1}
\end{equation}
where $M_P=2.4\times 10^{18}$\,GeV is the reduced Planck mass, 
$m_{3/2}$ is the gravitino mass, 
and $|U_{\widetilde{\gamma}\nu}|^{2}$ determines the neutrino content of the photino:
\begin{equation}
|U_{\widetilde{\gamma}\nu}|^{2}=\sum_{i=1}^{3}|N_{i1}\cos\theta_{W}+N_{i2}\sin\theta_{W}|^{2}.
\end{equation}
Here $N_{i1}$ ($N_{i2}$) is the Bino (Wino) component of the $i$-th neutrino, and $\theta_{W}$ is the weak mixing angle.
The same result for the decay width holds for the conjugated 
processes $\Psi_{3/2}\rightarrow\gamma\bar{\nu}_i$.

Assuming that this is the only relevant decay channel of the gravitino, its lifetime can then be written as 
\begin{equation}
{\tau}
(\Psi_{3/2}\rightarrow\sum_i\gamma\nu_i
)
=\frac{1}{2\Gamma\left(\Psi_{3/2}\rightarrow\sum_i\gamma\nu_i\right)}
\simeq 3.8\times 10^{27}\, {s}
\left(\frac{10^{-16}}{|U_{\widetilde{\gamma}\nu}|^2}\right)
\left(\frac{10\, \mathrm{GeV}}{m_{3/2}}\right)^{3}\ ,
\label{lifetimegamma}
\end{equation}
where the factor 2 takes into account the charged conjugated final states.
If $|U_{\widetilde{\gamma}\nu}|^{2}$ is small enough,
the gravitino can be very long lived 
compared to the current age of the Universe which is about $4\times 10^{17}$\,s.

We can easily estimate the value of $|U_{\widetilde{\gamma}\nu}|^{2}$
in the $\mu\nu$SSM~\cite{Choi:2009ng}. Using the mass insertion technique,
from the entries in the neutral fermion mass matrix (\ref{mixing3})
and Fig.~\ref{decay},
we can deduce that the relevant coupling for the mixing between the photino and the neutrinos is given approximately by $g_{1}v_{\nu}$, and as a consequence
$|U_{\widetilde{\gamma}\nu}|\sim |\frac{g_{1}v_{\tilde\nu}}{M_{1}}| \sim 10^{-6}$-$10^{-8}$, 
giving rise to
\begin{equation}
10^{-16} \lesssim |U_{\widetilde{\gamma}\nu}|^{2} \lesssim 10^{-12}.
\label{representative}
\end{equation}
One can confirm this estimation performing a scan of the low-energy parameter space of the $\mu\nu$SSM with the exact formulas above~\cite{Choi:2009ng}, imposing that neutrino data must be reproduced.
As a result of the scan, typically the mass of the neutralino is above 20\,GeV, and since $m_{3/2}$ is constrained to be smaller than that value, as we will see, the gravitino can 
safely be used as the LSP. Let us remark then, that 
under this assumption of gravitino DM, each candidate for LSP mentioned in the previous section would in fact be the next-to-LSP (NLSP), since the gravitino would be the LSP.
Nevertheless, the analysis of the phenomenology at the LHC would not be altered since the NLSP would also decay
into ordinary particles using the same channels as if it were the LSP.
Thus our analysis there can be applied exactly the same for the case of  neutralino/sneutrino/stau/squark/chargino NLSP with the gravitino as the LSP.

On the other hand, for the gravitino to be a good DM candidate we still need to check that it can be present in the right amount to explain the observed relic density $\Omega_{DM} h^2 \simeq 0.1$.
With the introduction of inflation, the primordial gravitinos are diluted
during the exponential expansion of the Universe. Nevertheless, after inflation,
in the reheating process, the gravitinos are reproduced again from the
relativistic particles in the thermal bath. The yield of gravitinos
from the scatterings is proportional to the reheating temperature, $T_R$, and
estimated to be~\cite{Bolz:2000fu}
\begin{equation}
\Omega_{3/2} h^2 \simeq 0.27\left(\frac{T_R}{10^{10}\ GeV} \right)
\left(\frac{100\ GeV}{m_{3/2}} \right)\left(\frac{M_{\tilde g}}{1\ TeV}\right)^2,
\label{oh2buchmuller}
\end{equation}
where $M_{\tilde g}$ is the gluino mass. As is well known, adjusting the
reheating temperature one can reproduce the correct relic density for each possible value of the gravitino mass.
For example for $m_{3/2}$ of the order of $1-1000$ GeV 
one obtains $\Omega_{3/2} h^2 \simeq 0.1$ for
$T_R\sim 10^8-10^{11}$ GeV, with $M_{\tilde g}\sim 1$ TeV.
Even with a high value of $T_R$ there is no cosmological gravitino problem, since the
NLSP decays to standard model particles much earlier than Big Bang nucleosynthesis (BBN) epoch
via $\rpv$ interactions.

Thus, the gravitino, which is a super-weakly interacting massive particle (superWIMP),
represents a good DM candidate. 
Most importantly, as pointed out in Ref.~\cite{Takayama:2000uz} for the case of $\rpv$, gravitino decays in the Milky Way halo would produce monochromatic gamma rays with an energy equal to half of the gravitino mass, and therefore its presence can, in principle, be 
inferred indirectly from gamma-ray observations. We will discuss this crucial issue in the next subsection.

\subsection{Detection 
}
\label{detection}

The detection of gravitino DM in several $\rpv$ scenarios has been studied 
in the literature 
considering the case of gravitinos emitting gamma rays when decaying in 
the smooth galactic halo and extragalactic regions at cosmological 
distances \cite{Takayama:2000uz,Buchmuller:2007ui,Bertone:2007aw,Ibarra:2007wg,Ishiwata:2009vx,Buchmuller:2009xv,Choi:2009ng,Restrepo:2011rj,Albert:2014hwa}, and
also in nearby extragalactic structures \cite{GomezVargas:2011ph}.
In the interesting case of the galactic halo, the gamma-ray signal is an anisotropic sharp line and the flux is given by
\begin{equation}
\label{halo}
\frac{d\Phi}{dE}(E)=\frac{\delta(E-\frac{m_{3/2}}{2})}{4 \pi \tau_{3/2} m_{3/2}}
\int_{\textrm{los}}\rho_{halo}(\vec{l})d\vec{l}\ .
\end{equation}
It is worth noting that this equation has two independent factors. The first one corresponds to the particle physics properties of the DM candidate. In particular, its lifetime, 
$\tau_{3/2}$, its mass, $m_{3/2}$, and a delta function associated to the fact that the gravitino decays into a photon (and a neutrino), producing therefore a line with an energy equal to $m_{3/2}/2$. 
The second factor corresponds to the astrophysics and is given by the integral 
along the line of sight $l$ of the halo DM density.

\begin{figure}[t]
 \begin{center}
 \includegraphics
[width=0.9\textwidth]{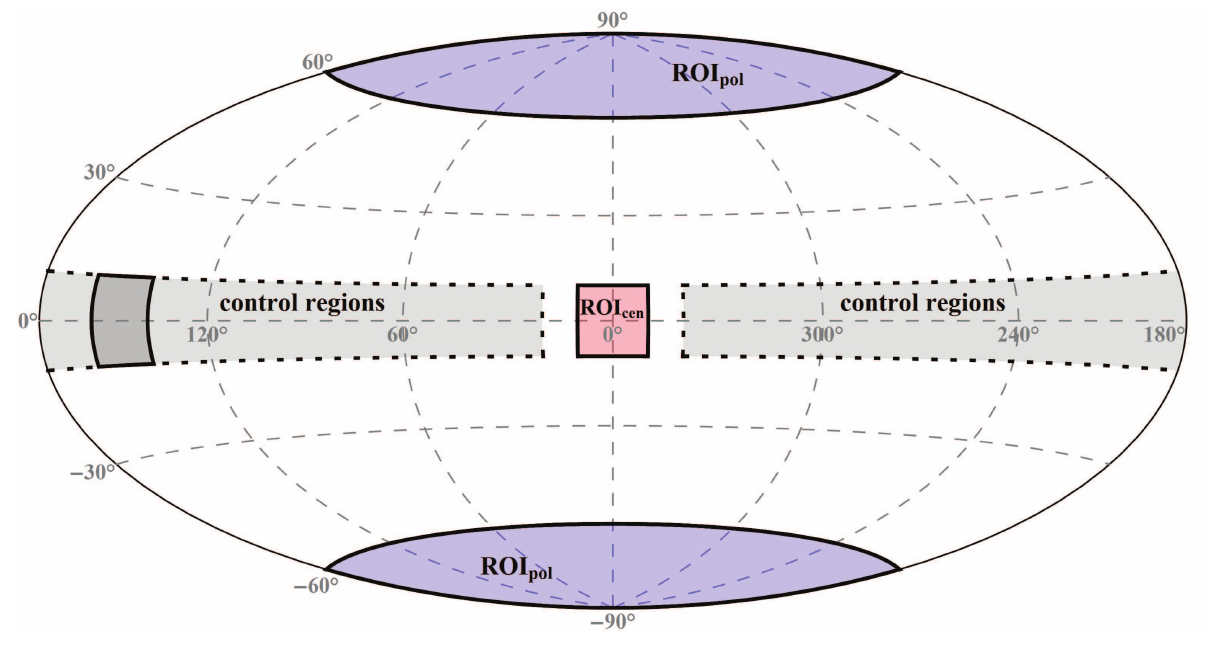}
\caption{
Skymap of the ROI used in our analysis~\cite{Albert:2014hwa}; plotted in Galactic coordinates
using the Hammer-Aitoff projection. The region ROI$_{pol}$ (blue) is optimized for the signal-to-background ratio in the case of DM decay, while the region ROI$_{cen}$ (red) is optimised for the signal-to-background ratio in the case of DM annihilation. The dashed line encloses the area for the control regions along the Galactic plane (light gray), while the gray region is an example of one of the 31 control regions used in the analysis.
}
    \label{fig:gfn}
 \end{center}
\end{figure}

A first analysis in the $\mu\nu$SSM of the possible detection of this kind of signal in the Fermi-LAT was carried out in Ref.~\cite{Choi:2009ng}.
Taking into account the data reported by Fermi at that time, from the non-observation of lines it was possible  to constrain the lifetime and the mass of the gravitino. In particular, the mass has to be around 10 GeV or smaller.
In a more recent work together with Fermi-LAT members~\cite{Albert:2014hwa}, a search for 100 MeV to 10 GeV gamma-ray lines was carried out using 62 months of Fermi-LAT data, and the implications for gravitino DM in the $\mu\nu$SSM were analyzed.
In this category 2 paper of the Fermi-LAT collaboration we used 
an Einasto profile with a finite central density~\cite{einasto,Navarro:2003ew}:
\begin{equation}
 \rho_{{Ein}}(r)=\rho_{\odot}\exp\left( -\frac{2}{\alpha}\left( \left( \frac{r}{r_s}\right) ^\alpha-\left( \frac{R_{\odot}}{r_s}\right) ^\alpha\right) \right) ,
\end{equation}
where we adopted $\alpha=0.17$ and $r_s=20\,$kpc for the case of the Milky Way and a local DM density of $\rho_{\odot}\simeq 0.4$ GeV cm$^{-3}$~\cite{Catena:2009mf,Weber:2009pt,Salucci:2010qr}.
Other halo profiles as well as uncertainties on the halo parameters were also taken into account, but all these profiles behave similar in the outer part of the Milky Way, where is our region of interest (ROI), and therefore the results are similar.
Concerning the ROI, we selected one that optimizes the signal-to-background ratio for searches for decay, where the Galactic poles are included, ROI$_{pol}: |b| > 60^o$. This is shown in 
Fig.~\ref{fig:gfn}

\begin{figure}[t]
 \begin{center}
 \includegraphics
[width=0.9\textwidth]{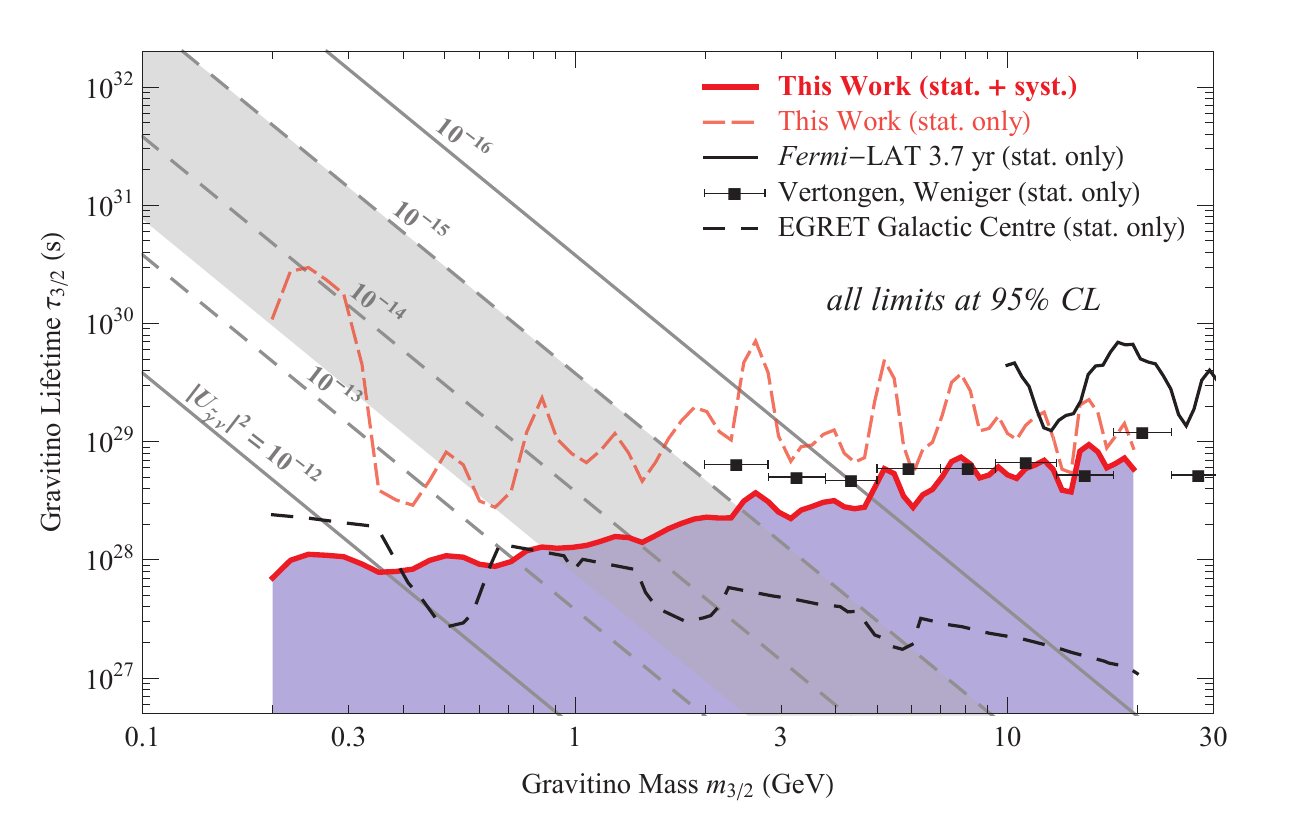}
\caption{Result of Ref.~\cite{Albert:2014hwa}, where the parameter space of decaying gravitino DM is given in terms of the gravitino lifetime and the gravitino mass. The diagonal band shows the allowed parameter space for gravitino DM in the $\mu\nu$SSM. The numbers on the solid and dashed lines show the corresponding value of the photino/neutrino mixing parameter, as discussed in section 4. The theoretically most favored region is colored in gray. We also show several 95\% CL lower limits on the gravitino lifetime coming from gamma-ray observations. The blue shaded region is excluded by the limits derived in the paper.
}
    \label{fig:results}
 \end{center}
\end{figure}

The final result of the analysis is shown in Fig.~\ref{fig:results}. We did not find any statistically significant spectral lines and have set robust limits on DM interactions that would produce monochromatic gamma rays.
When these limits are applied to the $\mu\nu$SSM, under the assumption that the gravitino is the DM, we find that the mass must be
$m_{3/2}<4.8$ GeV and the lifetime $\tau_{3/2}>7.9\times 10^{27}$ s at 95\% CL if we assume that all the DM in the Universe is in the form of gravitinos.

In a work in preparation~\cite{gravitino}, we are performing a deeper exploration of the 
$\mu\nu$SSM parameter space, taking also into account 3-body final states in the computation.
The preliminary result shows that in some regions of the parameter space is possible to increase the upper bound on the gravitino mass to about 20 GeV and to lower the lower bound on the lifetime to about $10^{25}$ s.

\section{Conclusions}
\label{conclusions}

The $\mu\nu$SSM solves the $\mu$ problem of SUSY models
and reproduces neutrino data, simply using couplings with the three families of RH neutrinos. These new couplings produce $\rpv$, generating a phenomenology very different from the one of the MSSM or the NMSSM. We have shown that
novel signatures of SUSY at the LHC are expected. 
In particular, all sparticles are potential candidates for the LSP, not only the usual lightest neutralino but also the lightest stau, squark, chargino, sneutrino.
The LSP is not stable leading to prompt or displaced vertices,
and producing final states with multi-leptons/taus/jets/photons and missing energy. 
On the other hand,
the gravitino turns out to be an interesting candidate for DM, since its lifetime 
is longer than the age of the Universe.
It can be searched through gamma-ray observations such as those of the Fermi-LAT. The non-observation of spectral lines allows to set robust limits on the parameters of the model. In particular, depending on the region of the parameter space, the gravitino mass must be smaller than about $5-20$ GeV
and the lifetime larger than about $10^{25-28}$ s.

\section*{Acknowledgments}
I gratefully acknowledge the local organizers of DSU 2015 for the fantastic atmosphere of the workshop, and the marvellous visit to Yukawa's office where I am sure that all participants hope to have gotten some inspiration.

This work was supported in part by the Spanish grant 
FPA2015-65929-P MINECO/FEDER UE, and by the Programme SEV-2012-0249 `Centro de Excelencia Severo Ochoa'. The author also acknowledges specially the support of the Spanish MINECO's Consolider-Ingenio 2010 Programme under grant MultiDark CSD2009-00064. 

\bibliography{munussm_v2}
\bibliographystyle{JHEP}

\end{document}